# Maturity assessment and maturity models in healthcare: A multivocal literature review

Ayça Kolukısa Tarhan
Department of Computer Engineering
Hacettepe University, Ankara, Turkey
atarhan@cs.hacettepe.edu.tr

Vahid Garousi
School of Electronics, Electrical
Engineering and Computer Science
Queen's University Belfast
United Kingdom
v.garousi@qub.ac.uk

Oktay Turetken
Department of Industrial Engineering
and Innovation Sciences
Eindhoven University of Technology
Eindhoven, the Netherlands
o.turetken@tue.nl

Mehmet Söylemez
Department of Computer Engineering
Hacettepe University, Ankara, Turkey
soylemez.mehmet@gmail.com

Sonia Garossi
Watford General Hospital
Herts, United Kingdom
sonia.garossi@whht.nhs.uk

**Abstract:**

*Context*: Maturity of practices and infrastructure in healthcare domain directly impacts the quality and efficiency of healthcare services. Therefore, various healthcare administrations (e.g., hospital management to nation-wide health authority) need to assess and improve their operational maturity. *Objective*: This study aims to review and classify studies that propose/use maturity assessment or maturity models (MMs) as a vehicle to achieve operational excellence in healthcare domain. *Method*: To achieve this objective, we performed a Multivocal Literature Review (MLR) that is a form of Systematic Review and includes data from the grey literature (e.g., white papers and online documents) in addition to formal, peer-reviewed literature. *Results*: Based on 101 sources, 80 of which are from the peer-reviewed literature and 21 are from the grey literature, we identified 68 different MMs on, e.g., telemedicine, care pathways, and digital imaging. We reviewed them with respect to various aspects including: types of research and contribution; list of MMs proposed/used with their subject focuses; elements of maturity/capability; and application scope or scale. In the synthesis of empirical benefits of using MMs, two were found significant: (1) Identifying issues and providing guidance for improvement in healthcare contexts; (2) Improving efficiency, effectiveness, performance, and productivity. *Conclusion*: This MLR provides an overview of the landscape and serves as an index to the vast body of knowledge in this area. Our review creates an opportunity to cope with the challenges in getting an overview of the state-of-the-art and practice, choosing the most suitable models, or developing new models with further specialties.

## 1 INTRODUCTION

Maturity of practices, operations and infrastructure in the healthcare domain is of high importance [1]. However, these qualities often vary among healthcare provider units (e.g., hospitals) and in some cases could be far from perfect. Establishing process thinking and achieving quality management is vital to assure service maturity, and to continually improve service maturity in such a complex, dynamic, and multidisciplinary domain [2, 3].

The notion of "maturity" was first proposed by Phillip Crosby [4] and is defined as "*the state of being complete, perfect, or ready*" [5]. In a more general view, a maturity model (MM) is a conceptual framework that consists of a sequence of discrete maturity levels for a class of processes in one or more business domains, and represents an anticipated, desired, or typical evolutionary path for these processes [6, 7]. Maturity models (also referred to as 'capability frameworks') have been widely used in many domains (such as industrial engineering, software engineering and information technology) for the purpose of process assessment and improvement [8-10].

Maturity models have been proposed also in the healthcare domain with the purpose to assess and improve the maturity of healthcare practices, operations and infrastructure [11, 12]. Some of these MMs focus on the specific sub-areas, such as telemedicine, care pathways, digital imaging, picture archiving and communication systems (PACS), and facilities management. Given the large body of knowledge in this area, there is a need to review the scope and characteristics of available models together with their use and usefulness. Healthcare practitioners, decision makers, and researchers can then eliminate "reinventing the wheel" to get the state-of-the-art and –practice, choose the "right" models that are most suitable for their needs, or develop new models with further specialties. Therefore, this study is aimed at providing insight into and creating awareness on the maturity models proposed as a vehicle to achieve operational excellence in the healthcare domain.

To achieve the abovementioned objective, we performed a comprehensive literature review on MMs proposed and/or used in the healthcare domain. We used Multivocal Literature Review (MLR) [13] as the research method, which is a form of a systematic literature review (SLR) and includes grey literature (e.g., blog posts, white papers, presentations) in addition to published (formal) academic literature. While SLR is a common method used to conduct a literature review on the published (peer-reviewed) literature on a given topic, the significance of including the grey literature in addition to academic one has been emphasized in the healthcare community, e.g., [14-16].

We used systematic guidelines [13, 17] to search for the sources, categorize and synthesize them with respect to a number of review questions (RQs). We searched the academic literature using the Google Scholar, PubMed, and ScienceDirect; and the grey literature using the regular Google search engine. In the remainder of this paper, we use the term "source" to refer to a formally-published paper/article, or a resource from the grey literature. Through the RQs, we addressed the following properties of the sources: (1) types of research and contribution, (2) newly proposed vs. existing MM used in the studies; (3) the aspects of maturity/capability addressed by the MMs (technology, business process, people, etc.); and (4) the application scope or scale of the MMs (e.g., single departments of a single hospital, multiple hospitals in a city, region or province). More specifically, the main contributions of this article can be summarized as follows:

- An overview of the state-of-the-art and –practice of 101 sources that propose or use MMs in the healthcare domain;
- A list of 68 MMs proposed or used in this domain, with details of selected examples;
- A classification scheme of the sources with respect to the attributes of: research type, contribution type, subject focus, contained aspect, applied scope, scale of empirical evidence, and reported benefits of the MMs; and
- A classification and synthesis of the sources with respect to the attributes mentioned above.

The remainder of this paper is structured as follows. Related work is presented in Section 2. We describe the research method and the review planning in Section 3. Section 4 presents the results of the literature review. In Section 5, we discuss our observations and suggestions, and potential threats to the validity of this MLR. Finally, in Section 6, we draw conclusions and suggestions for future work.

## 2 RELATED WORK

Since this paper is an MLR on maturity models in healthcare, the related works are the existing *secondary* studies on MMs in healthcare. A secondary study is a study of regular (primary) studies.

By a literature search, we were able to identify 12 secondary studies on MMs in healthcare, as listed in Table 1. Secondary studies are usually of four types: Systematic mapping (SM) studies, SLRs, MLRs, and regular reviews (surveys). For each paper in Table 1, we also provide the title of the paper to show its scope, its type (SLR, SM, and regular review, etc.) and the number of papers reviewed by the study.

Our current MLR differs from the existing works in that it is the first comprehensive review on maturity assessment and maturity models in healthcare, which covers both the formal and the grey literature and clusters subject focuses and aspects studied so far. In terms of number of papers reviewed by a secondary study, it is seen from Table 1 that our MLR reviews also the highest number of papers (101 sources) compared to the existing secondary studies and derives a broad classification scheme.

## 3 RESEARCH METHOD

We performed our study based on the guidelines for conducting SLR/MLR studies provided in the healthcare domain, e.g., [28, 29], and also in other domains, e.g., [13, 17].

The area of healthcare MMs is cross-disciplinary in nature, as it includes various stakeholders from health services and hospital management, to information systems (IS) and information technology (IT), each on the side of academia or practice. Furthermore, we found that some models in this area have been presented in grey literature and not formally published. Conducting a typical SLR without including the grey literature would not have included many important and useful MMs that have emerged from practice and is available in the grey literature. Therefore, we followed a natural approach by including the grey literature in the review and conducted an MLR. This decision is also in alignment with the literature considering the use of MLRs as a means to close "*the gap between academic research and professional practice*" [13].

We discuss various aspect of our research method next.

Table 1- Secondary studies on MMs in healthcare (order by year of publication)

| Reference | Year of pub. | Title | Type of study | # Primary studies (papers) reviewed |
|---|---|---|---|---|
| [18] | 2009 | A PACS maturity model: A systematic meta-analytic review on maturation and evolvability of PACS in the hospital enterprise | SLR | 34 |
| [19] | 2011 | Evolution of information systems and technologies (IST) maturity in healthcare<br>• As part this paper, the authors provided a survey of MMs focusing on the IST management in healthcare | A brief regular review (survey) as part of the paper | 5 |
| [20] | 2013 | Composite quality of care scores, electronic health record maturity models, and their associations; preliminary literature review results | SLR | 53 |
| [21] | 2013 | Towards a business intelligence (BI) maturity model for healthcare<br>• Presented a review of existing BI MMs to determine their adequacy for use in healthcare. | Regular review | 15 |
| [22] | 2015 | Quality system maturity model for medical devices-medical device innovation consortium<br>• Provided an overview of various MMs in the medical device industry and how they have been implemented and leveraged. | Regular review | 22 |
| [23] | 2016 | A patient-centered framework for evaluating digital maturity of health services: A systematic review | SLR | 28 |
| [24] | 2016 | Information systems and technologies maturity models for healthcare: A systematic literature review | SLR | 14 |
| [25] | 2016 | Maturity models for hospital information systems management: Are they mature? | Regular review | 3 |
| [11] | 2016 | Maturity models of healthcare information systems and technologies: A literature review | SLR | 14 |
| [12] | 2016 | The use of maturity/capability frameworks for healthcare process assessment and improvement | SM/SLR | 29 |
| [26] | 2017 | A review and comparison of maturity/capability frameworks for healthcare process assessment and improvement | SLR | 6 |
| [27] | 2017 | A maturity model for hospital information systems | A brief SLR as part of the study | 14 |
| This study | 2019 | Maturity assessment and maturity models in healthcare: a multi-vocal literature review | MLR | 101 |

### 3.1 GOAL AND REVIEW QUESTIONS

The goal of this study was to review and classify studies that propose/use maturity assessment or maturity models in healthcare domain. Based on this goal, we raised the following review questions (RQs):

- **RQ 1- Contribution and research types:**
  - **RQ 1.1- Contribution types:** How many sources have presented new MMs, methods/techniques, tools, metrics, or processes in this area?
  - **RQ 1.2- Research types:** What types of research methods were used by authors of the sources? Some sources only propose solutions without extensive validations, while some other sources present in-depth evaluation of their approach (e.g., rigorous empirical studies).
- **RQ 2- Properties of the proposed MMs:**
  - **RQ 2.1- MMs in this area:** What models have been proposed in this area? What are their subject focuses?
  - **RQ 2.2- Aspects of maturity/capability framework:** What aspects (e.g., business process, organizational, technology, and people) does each of the proposed MM address?
  - **RQ 2.3- Applicability scope/scale:** What is the applicability scope/scale of the MMs? The models could be applied within the scope of, for example, single/multiple department(s), or single/multiple hospital(s). In this RQ, we consider the expected/possible applicability of the models, and not necessarily the scale of the actual empirical study presented in a given source.
- **RQ 3- Properties of the empirical studies:**
  - **RQ 3.1- Scale of empirical study:** In what scales the empirical studies have been conducted, e.g., the number of hospitals that the proposed MM has been applied in? As different from RQ 2.3, here we consider the "actual" reported empirical results presented in a source.
  - **RQ 3.2- Reported benefits of MM applications:** What benefits (quantitative or qualitative) have been reported as a result of applying the models? The application of such a model is sometimes claimed to be costly, thus enough benefits should be provided to ensure justifying the associated costs.

## 3.2 MULTIVOCAL REVIEW PROCESS

Figure 1 outlines the process that lies at the basis of this study, which consists of three phases:

- Source selection (Sections 3.3, and 3.4)
- Development of the classification scheme (Section 3.5)
- Systematic classification, synthesis and review (Section 4)

The process starts with the selection of sources in the academic and grey literature. Next, a classification scheme (map) is developed following a structured approach. The map is then used to conduct systematic classification, which is followed by the synthesis and reporting of the results.

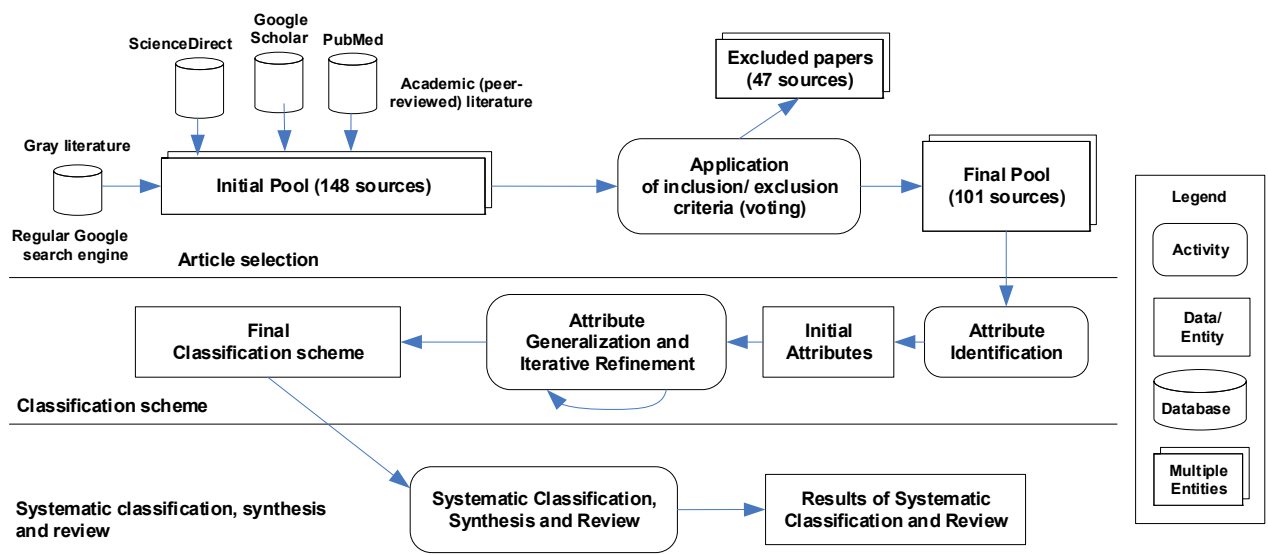

**Figure 1- An overview of MLR process**

## 3.3 SEARCH AND SELECTION OF SOURCES

To search for formally-published (peer-reviewed) papers, we used the Google Scholar, PubMed and ScienceDirect search engines. There have been numerous studies, e.g., [30-34] comparing the effectiveness, strengths and weaknesses of these academic search engines. We carefully reviewed these studies to be as objective and systematic as possible in our choice of search engines.

We also conducted a number of experimental search activities for papers in this area. We observed that the topic of MMs in healthcare is a cross-disciplinary subject and that many papers in this area (e.g., Sources #10, 17, and 24 in the pool of studies [35]) are published in venues that address the domains outside of the "core" healthcare (e.g., management, information technology, and computer science), and thus are not covered (could not be found) by PubMed. Since Google Scholar had a much larger index (search space) than PubMed, we first used Google Scholar and then PubMed to populate the candidate pool of sources.

Through an iterative process, we developed the search string of this MLR study as: *(maturity model OR maturity assessment) AND (health care OR healthcare OR health)*. We should note that the search process and source selection of this study was conducted in March 2018, thus sources published by that time had a chance to be included in our pool.

We performed the searches in our list of search engines (i.e. Google Scholar, PubMed and ScienceDirect) using the above keywords. Via Google Scholar, we found 76 candidate formally-published papers. Via PubMed, we found four additional candidate formally-published papers which were not found through Google Scholar. They were Sources #13, 25, 80 and 90. Full details can be found in the study's transparent online spreadsheet: https://goo.gl/Z4jxBd. Finally, we searched in the ScienceDirect digital library but did not find any additional papers.

To search for grey literature sources, we used the Google search engine that provides comprehensive access to grey literature [31]. This popular web search engine (also the most widely used one with respect to [36]) has been used in many MLR studies in the past, e.g., in health [37] and management sciences [38].

As a result of the first search phase, we obtained an initial pool of 148 sources on which we applied the criteria for inclusion/exclusion and quality assessment.

## 3.4 CRITERIA FOR INCLUSION/EXCLUSION AND QUALITY ASSESSMENT OF SOURCES

We carefully defined the inclusion and exclusion criteria to ensure including all relevant sources but not those which were out of scope. Our inclusion criteria were:

- Sources that proposed or used maturity models in the healthcare domain.
- Sources which are directly related to maturity assessment in the healthcare domain.
- Sources that are in English and their full-texts were accessible via our institutions' subscriptions or were freely available on the Internet. (Please note that one author of this article has medical affiliation and we could have good-enough access to healthcare publication venues.)

In addition to the criteria above, we also defined and applied criteria for checking quality of the sources. Quality assessment is specifically focused on determining the sources that report sufficient information to compare other studies in terms of answering defined review questions [39]. Therefore, we developed and used the following quality assessment criteria to ensure that sufficient information is reported in the sources:

- Is there a clear statement that the study proposes/uses an MM for healthcare?
- Can it be inferred from the study that the MM is newly proposed, or it is only applied as already available?
- Can the organizational aspects that the MM involves, e.g., process, technology, and people, be extracted from the study?
- Can the scope/scale in which the MM could be applied be extracted from the study?

All sources were examined in detail by the first three researchers in the *voting stage* with respect to the inclusion/exclusion and quality assessment criteria given above, and the sources which did not meet these criteria were excluded. Each of the first three researchers did independent voting on either to include or exclude each source. In the case of disagreements, discussions were conducted and the fifth author who is a medical practitioner (pathologist) was consulted to reach a consensus.

As previously stated, search process and source selection of this study was conducted in March 2018. Only a single 2018 paper was found [40], which was excluded in order not to provide a partial view for 2018. As a result, the time window for the pool sources was set to the yearly range of 1996-2017. At the end of applying the inclusion/exclusion and quality criteria, 47 sources were excluded (see Figure 1), and our review pool was finalized with 101 sources.

## 3.5 DEVELOPMENT OF THE CLASSIFICATION SCHEME AND DATA-EXTRACTION

To conduct a structured and systematic classification of the sources, there is a need for a classification scheme (systematic map) [28, 29]. We developed our classification scheme through an iterative process of analyzing the sources in the pool, identifying relevant list of attributes, and refining them to derive the final map. Our goal was to categorize the sources to build a complete picture of the research area. Driven by our review questions, we identified the key attributes of the sources and refined them into classification scheme using an iterative approach. Table 2 shows the classification scheme that we developed after applying the process described above.

Taking our RQs and the classification scheme as the basis, we thoroughly reviewed the sources in our pool. In tagging each study based on the classification scheme, we incorporated as much explicit "traceability" links between our mapping and the primary sources as possible. That is, we added comments to the cells of our mapping matrix by verbatim copy/paste of text from the source acting as the "traceability" link. Such comments facilitated peer reviewing (that we conducted among the authors) and are also useful for other researchers reviewing the resulting paper pool and the supplementary data.

As in the initial voting stage, during the detailed analysis phase, three researchers extracted and analyzed data from the portion of the sources, and then peer reviewed each other's extracted data. In the case of disagreements among the researchers, issues were resolved by holding discussions with all authors to ensure quality and validity of the data and their relevance in the healthcare domain.

**Table 2- Classification scheme (systematic map) developed and used in this study**

| RQ | | Attribute/Aspect | Categories | Multiple (M) / Single (S) Value |
|---|---|---|---|---|
| RQ 1 | 1.1 | Contribution type | New (maturity) model, method/technique, tool, metric, | M |

| RQ | | Attribute/Aspect | Categories | Multiple (M) / Single (S) Value |
|---|---|---|---|---|
| | | | process, empirical study only, other | |
| | 1.2 | Research type | Validation research (weak empirical study), evaluation research (strong empirical study), solution proposal, philosophical paper, experience paper, opinion paper, other | S |
| RQ 2 | 2.1 | MMs proposed with subject focuses | Name of the MM proposed in the source, with its subject focus(es) | S (name) / M (subject focus) |
| | 2.2 | Aspects covered by the MM | Business process, technology, people, other | M |
| | 2.3 | Scope of MM application | Single departments of a hospital/healthcare institution; multiple departments of a hospital; single healthcare institution; multiple healthcare institutions; city, region or province; government (nation-wide); international; other | M |
| RQ 3 | 3.1 | Scale of the empirical study | Number of hospitals/healthcare institutions, any other size/scope related metric (e.g., number of experts, number of survey or interview participants): numerical values | M |
| | 3.2 | Reported benefits of MM application | Quantitative benefits, qualitative benefits: open text | M |

## 4 RESULTS

Our final pool included a total of 101 sources, from which 80 sources were from the formal literature and 21 were from the grey literature. For a complete list of these sources, please refer to [35]. Below we report a summary on the trends in the final pool of sources, which is followed by the results in Sections 4.1 through 4.3 with respect to our RQs.

Figure 2 shows the cumulative number of sources per year by literature type (formally published versus grey literature). We can see that the sources in both literature categories have been on a steady increasing trend. While many scientists are publishing their works in this area, many practitioners/professionals are using the online platforms such as blogs for posting their works in this area as grey literature. We can see from the figure that the topic has received an active attention in recent years.

Out of 80 sources from the formal literature, 60 were journal papers, 13 were conference papers, 4 were theses, 2 were book chapters, and 1 was a workshop paper. Out of 21 sources from the grey literature, there were 13 white papers, 4 presentations, 3 web pages, and 1 blog post. Journal papers form the majority (69%) in the pool, denoting a considerable level of maturity in the research field.

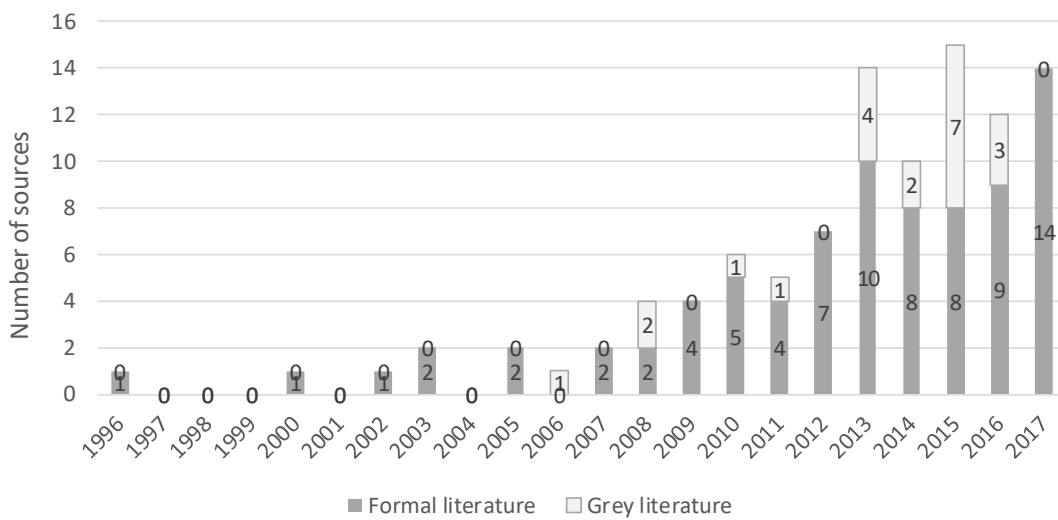

**Figure 2- Number of sources per year**

## 4.1 RQ 1– CONTRIBUTION AND RESEARCH TYPES

### 4.1.1 RQ 1.1– Contribution types

Figure 3 shows the annual cumulative breakdown of sources by contribution types. We also provide, under the figure, the references of the sources for each type. Please note that a source may contribute more than one contribution type. The top three contribution facets are new MMs, empirical (case) studies, and methods/techniques which have appeared in 68, 26 and 4 sources, respectively.

We can see that across different years; different contribution facets have been studied. The general observable trend is that new MMs are continuously being offered in this area. In addition, the trend with sources that have major contributions as empirical (case) studies is a promising sign of practice by the community.

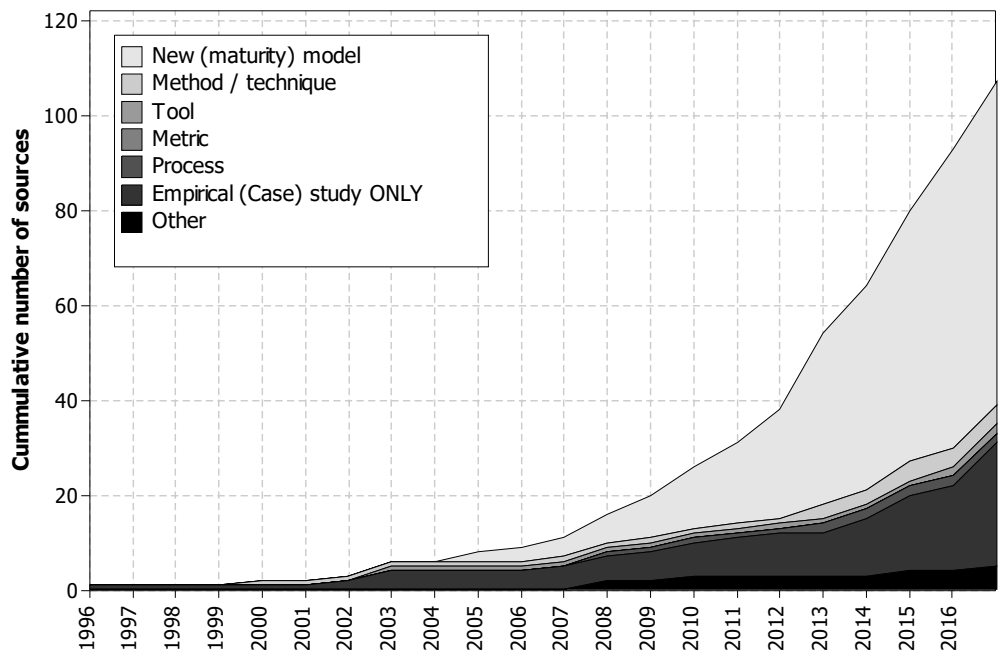

| Contribution facet types | # of sources | References |
|---|---|---|
| New (maturity) model | 68 | Too many to be listed. Refer to the online spreadsheet (https://goo.gl/ENscIn) |
| Method/technique | 4 | [Sources 2, 22, 59, 74] |
| Tool | 2 | [Sources 54, 64] |
| Metric | 0 | - |
| Process | 2 | [Sources 29, 39] |
| Empirical (case) study only | 26 | [Sources 12, 17, 18, 19, 20, 25, 26, 36, 40, 42, 48, 49, 53, 57, 58, 61, 64, 75, 79, 83, 84, 86, 93, 94, 95, 97] |
| Other | 5 | [Sources 10, 16, 28, 29, 34] |

**Figure 3- Sources by contribution type**

In Figure 4, we present a bird's-eye view of MMs by showing a word-cloud of the MM names (Section 4.2.1 discusses a review of these MMs). As we can see, there have been many MMs focusing on the topics of information technology, electronic systems, and telemedicine. These topics highlight the importance of increasing digitalization in this domain.

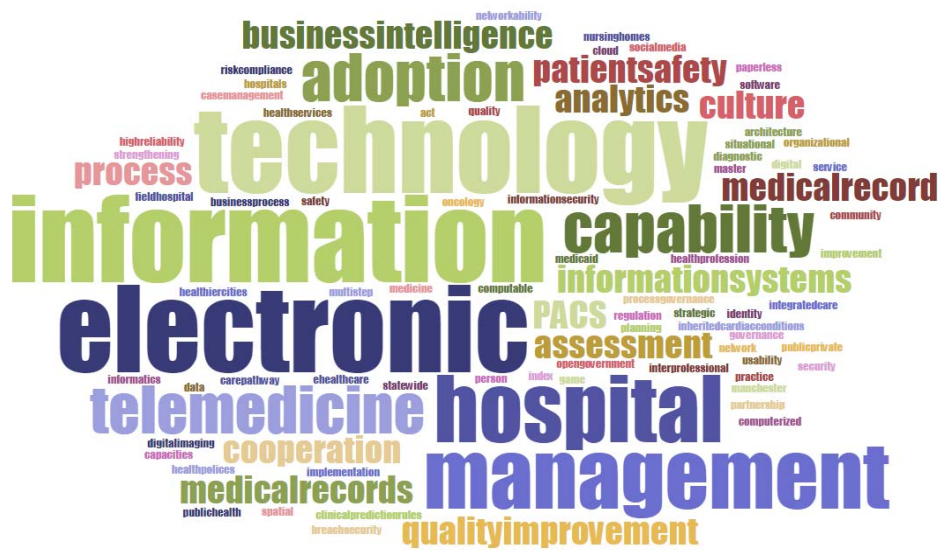

**Figure 4- A word-cloud of the MM names presented in the literature**

Four sources contributed methods and frameworks in this area. [Source 2] presents a framework for a domain-specific business intelligence MM and applied it in healthcare. [Source 22] developed a maturity assessment tool and technique in the subject of clinical governance. [Source 59] proposed a maturity model for identity management and a procedure for applying the MM, and reports from practice on its successful use in two large Swiss hospitals. [Source 74] proposed a method to identify selected number of items from a previously validated Quality Improvement (QI) Maturity Tool as the basis for calculating organizational and system-level QI maturity scores in time.

As mentioned above, the contributions of 26 sources were empirical (case) results only. For example, the research in [Source 12] sought further empirical evidence of IS and IT maturity in the context of UK's National Health Service (NHS) acute Trust hospitals. A survey was used to collect data from over 70 top and middle managers representing four Trust hospitals in the UK. Statistical analysis of these data provided evidence that six of 23 maturity characteristics identified in existing models can currently be used to differentiate the maturity of NHS acute Trust hospitals. The aim of the study in [Source 17] was to investigate the challenges of assessing process maturity in healthcare institutions using a generic business process maturity model, i.e., OMG's Business Process Maturity Model (BPMM), and to explore the opportunities for future work that would facilitate process maturity assessment and improvement in the healthcare domain. [Source 20] reported an empirical audit of a health agency for preparedness with respect to the implementation of e-health management, using the COBIT framework (Control Objectives for Information and Related Technology) [41]. Section 4.3 provides further insights into the pool of empirical studies in response to RQ 3.

Two sources offered tools as one of their main contributions. In [Source 54], the proposed healthcare quality maturity assessment model was incorporated in an interactive Excel worksheet that visually displayed the quality maturity-level risk meters, which we considered under the 'tool' category. Another tool-presenting work was [Source 64], which presented a software tool assisting the users in applying the *General Practice Information Maturity Model (GPIMM)*.

Two sources contributed processes. [Source 29] developed a Capability Maturity Model (CMM)-based security risk assessment 'process' for patient-centered healthcare systems, and applied it in the context of Personal Health Records (PHR). The Capability Maturity Model (CMM) [10] is a framework developed based on the data collected from organizations that contracted with the U.S. Department of Defense. The term 'maturity' in CMM relates to the degree of formality and optimization of processes, from ad-hoc practices, to formally defined steps, to managed result metrics, to active optimization of the processes. CMM's aim is to improve software development processes, but it has also been applied in other domains. [Source 39] proposes a process for maturity assessment based on the MM introduced in that paper, i.e., the Healthcare IT Maturity Model.

Five sources focused on 'other' types of contributions. [Source 10] proposed a software process improvement (SPI) model for the UK's healthcare industry by using organizational maturity (based on ISO/IEC 15504 [9]) as the guideline. [Source 16] discussed authors' experiences on applying the CMMI-SVC (CMMI for Services) model to healthcare services, specifically for in-hospital pharmaceutical and respiratory services. Similarly, [Source 28] shared the encountered experience and challenges that the authors faced during the design and implementation of three MMs for distinct improvement purposes in hospitals. [Source 29] mapped the components of the CMM into the regulations of the American Health Insurance Portability and Accountability Act (HIPAA). [Source 34] discussed opinions on using Digital Maturity Assessment (DMA) in order to understand the level of readiness for using technology in UK's National Health Services.

### 4.1.2 RQ 1.2– Research types

Based on the classification scheme described in Section 3.5, we classified the sources into seven types of research methods. Figure 5 shows the classification of the sources according to the type of research method they have followed. Recall that, for the research facet type, each study could be classified into only one category.

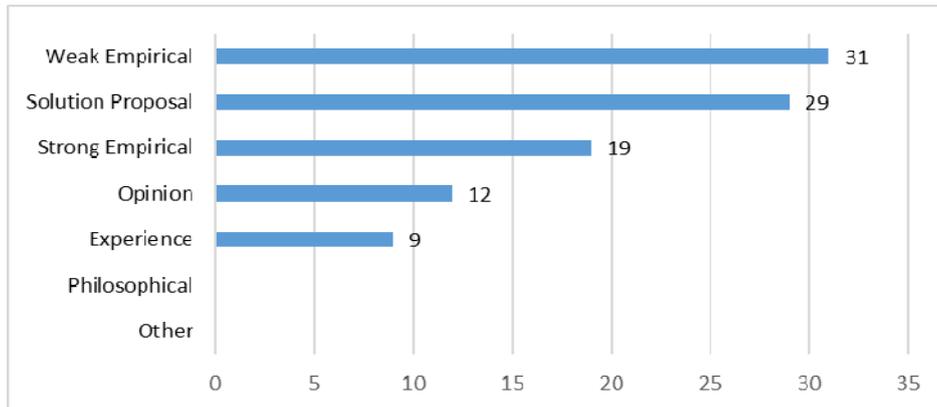

**Figure 5- Sources by type of research method**

A large ratio of sources falls under 'solution proposals' (with 29 sources), and weak and strong empirical studies (with 31 and 19 sources, respectively), which is a positive sign of empiricism in this community. Twelve sources reported 'opinions'. There are nine experience reports based on 'experience' of their authors. No source is classified as a philosophical paper.

## 4.2 RQ 2– VARIOUS PROPERTIES OF MMs

### 4.2.1 RQ 2.1– The proposed MMs with their subject focuses

We compiled the list of all the proposed MMs and gathered them in Table 3. In total, 68 of the sources proposed 68 new models in this area. As it can be seen from the table, there is a vast diversity on the topic of MMs in this area. Thus, we also extracted subject-focus of these models and provide an index of these subjects, as addressed by the sources, in alphabetical order in an online resource [35]. The most frequently concerned subjects by the models include: health services and information technology (n=6 each); public health, quality improvement, and safety culture (n=4 each); business intelligence, data analytics, e-health, electronic medical records (EMR), EMR adoption, hospital information system, integrated care, patient safety, security, and tele-medicine (n=3 each).

**Table 3- List of proposed MMs by the sources**

| # | Name of MM | Source # in online pool |
|---|---|---|
| 1 | Statewide master person index | 1 |
| 2 | Business intelligence (BI) maturity model | 2 |
| 3 | Business process model | 3 |
| 4 | Care pathway maturity model | 4 |
| 5 | Maturity model for hospital information systems | 5 |
| 6 | Maturity model for telemedicine implementation | 6 |
| 7 | Multistep maturity model for Electronic and Computable Diagnostic Clinical Prediction Rules (eCPRs) | 7 |
| 8 | Picture archive and communication system maturity model | 8 |
| 9 | Digital maturity of health services | 9 |
| 10 | Act on oncology model | 11 |
| 11 | A process maturity model for governance | 13 |
| 12 | Inter-professional practice capability framework | 14 |
| 13 | Open government maturity model | 15 |
| 14 | Case management maturity model | 21 |
| 15 | Cloud maturity model | 23 |

| 16 | Business intelligence maturity in healthcare | 24 |
|---|---|---|
| 17 | Analytics assessment maturity model | 27 |
| 18 | Health profession regulation strengthening framework | 30 |
| 19 | Telemedicine maturity model | 31 |
| 20 | Quality improvement maturity index | 32 |
| 21 | Digital imaging adoption model | 33 |
| 22 | E-healthcare maturity model | 35 |
| 23 | Electronic healthcare maturity model | 37 |
| 24 | Healthier cities maturity model | 38 |
| 25 | Healthcare IT maturity model | 39 |
| 26 | Inherited-cardiac-conditions (ICC) maturity model | 41 |
| 27 | A PACS maturity model for strategic situational planning | 43 |
| 28 | Game maturity model | 44 |
| 29 | Data management maturity | 45 |
| 30 | Hospitals cooperation maturity model | 46 |
| 31 | Healthcare network maturity model | 47 |
| 32 | Healthcare analytics maturity model | 50 |
| 33 | Healthcare breach security maturity model | 51 |
| 34 | Healthcare information security adoption model | 52 |
| 35 | Healthcare quality maturity model | 54 |
| 36 | High-reliability healthcare maturity model: | 55 |
| 37 | Hospital medicine maturity model | 56 |
| 38 | Identity management maturity model | 59 |
| 39 | Informatics capability maturity model | 60 |
| 40 | IT capacities maturity model | 62 |
| 41 | Governance, risk and compliance maturity model | 63 |
| 42 | A maturity model for Hospital Information Systems (HIS) management | 65 |
| 43 | Field Hospital Maturity Model (FHMM) | 66 |
| 44 | Maturity model for integrated care | 67 |
| 45 | Electronic medical record (EMR) adoption framework | 68 |
| 46 | B3-Maturity Model (B3-MM) | 69 |
| 47 | Use of electronic medical records (EMR) maturity model | 70 |
| 48 | Quality Improvement (QI) maturity tool | 71 |
| 49 | Health in all polices maturity model | 72 |
| 50 | Medicaid information-technology architecture maturity model | 73 |
| 51 | Patient safety culture improvement tool | 76 |
| 52 | Manchester patient safety framework | 77 |
| 53 | Process management maturity model | 78 |
| 54 | Electronic medical record maturity model (EMR-MM), | 80 |
| 55 | Healthcare usability maturity model | 81 |
| 56 | Organizational public-private partnership maturity model | 82 |
| 57 | Safety culture assessment in community | 85 |
| 58 | Social media maturity model | 87 |
| 59 | Spatial maturity in healthcare | 88 |
| 60 | Software process capability/maturity models for the asynchronous store-and-forward telemedicine systems | 89 |
| 61 | Computerized medical records (CMR) maturity model | 90 |
| 62 | Healthcare paperless maturity model | 91 |
| 63 | Healthcare security maturity model | 92 |
| 64 | Public health it maturity | 96 |
| 65 | IT in nursing homes | 98 |
| 66 | Telemedicine service maturity model | 99 |
| 67 | Hospital cooperation maturity model | 100 |
| 68 | Networkability maturity model | 101 |

**4.2.2 RQ 2.2– Aspects covered by the MMs**

We classified the aspects covered by the proposed MMs. The possible aspects (categories of classification) were process, technology, people, and other aspects. Figure 6 shows the distribution of the sources with respect to these aspects. Note that, the MM presented in a given a source may cover more than one single aspect, e.g., [Sources 1 and 2], thus the sum of the four categories in the figure is greater than number of sources (101).

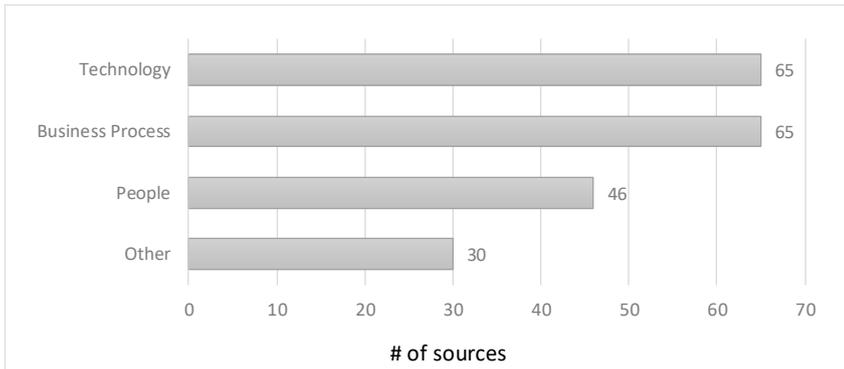

*Technology*: [Sources 1, 2, 3, 4, 5, 6, 8, 9, 10, 11, 12, 19, 21, 23, 24, 27, 29, 31, 33, 34, 35, 36, 37, 38, 39, 40, 43, 44, 46, 48, 49, 51, 52, 53, 54, 55, 56, 57, 58, 59, 60, 62, 65, 67, 68, 69, 70, 73, 78, 79, 81, 84, 88, 89, 90, 91, 91, 93, 94, 95, 96, 98, 99, 100, 101]

*Business Process*: [Sources 1, 2, 4, 6, 8, 10, 11, 12, 15, 16, 17, 18, 19, 20, 22, 24, 25, 26, 30, 31, 32, 34, 38, 40, 41, 43, 44, 45, 46, 47, 51, 52, 54, 55, 56, 58, 59, 60, 64, 66, 67, 69, 71, 72, 73, 74, 75, 76, 77, 78, 79, 80, 82, 83, 84, 85, 86, 89, 90, 91, 93, 94, 98, 99, 100]

*People*: [Sources 2, 3, 4, 5, 6, 9, 10, 12, 13, 19, 22, 31, 38, 41, 42, 44, 46, 51, 54, 55, 56, 57, 59, 60, 64, 65, 67, 69, 71, 74, 75, 76, 77, 78, 79, 80, 83, 84, 85, 86, 87, 89, 96, 98, 99, 100]

*Other*: [Sources 1, 2, 3, 7, 14, 15, 19, 22, 24, 34, 36, 38, 45, 46, 50, 54, 58, 59, 63, 65, 66, 69, 75, 78, 79, 83, 86, 87, 97, 100]

**Figure 6- Mapping of sources with respect to the aspects of the proposed MMs**

Table 4 shows the levels in the staged representation (or dimensions in continuous representation [42, 43]) of example models in different aspect categories.

The "business process" aspect is covered in 65 sources. For example, [Source #15] proposed an Open Government Maturity Model for participation/collaboration processes between public and government agencies. [Source #30] developed a framework to measure maturity of regulations in healthcare, and [Source #40] presented an MM for enterprise architecture and applied it in the context of Veterans health administration.

The "technology" aspect of maturity is also covered in 65 sources. As an example, [Source #5] defined an MM to assess hospital information systems. [Source #23] proposed a cloud maturity model for the healthcare industry, and [Source #33] described a digital imaging MM.

The "people" aspect of maturity is addressed in 46 sources. For example, [Source #19] defined an assessment tool for patient safety culture in Manchester. [Source #42] proposed a people capacity MM in medical record wards of Iranian hospitals. [Source #100] defined a cooperation MM for hospitals.

**Table 4- Three example MMs focusing on different aspects in healthcare**

| MM name | Source # | Aspect Category | Staged or continuous? | Levels (or dimensions) of the model |
|---|---|---|---|---|
| Open Government Maturity Model (OGMM) | Source #15 | Business process | Staged | • Level 1: initial conditions<br>• Level 2: data transparency<br>• Level 3: open participation<br>• Level 4: open collaboration<br>• Level 5: ubiquitous engagement |
| Cloud maturity model for the healthcare industry | Source #23 | Technology | Staged | • Level 1: Departmental, niche applications<br>   o Medical imaging archiving<br>   o Personal health records or PHRs<br>   o Analytics<br>• Level 2: Core health IT systems<br>   o Electronic health records (EHR)/Health information exchange (HIE)<br>   o Scheduling/practice management<br>   o Clinical decision support<br>   o Quality reporting<br>• Level 3: Virtualized, integrated health networks<br>   o Health plans<br>   o Hospitals, clinics and labs<br>   o Pharmacies<br>   o Patients and caregivers<br>• Level 4: Seamless care delivery<br>   o Anywhere, anytime access<br>   o Personalized care plan<br>   o Real-time visibility (cost, quality) |
| People capacity maturity model (PCMM) | Source #42 | People | Continuous | • Dimension 1: Staffing process area<br>• Dimension 2: Communication and coordination process area<br>• Dimension 3: Work environment process area<br>• Dimension 4: Performance management process area<br>• Dimension 5: Training and development process area<br>• Dimension 6: Compensation process area |

Under the "other" category, 30 sources covered various aspects including work culture, strategy, governance, leadership, interoperability, and data. Readers are encouraged to use Figure 6 to trace the references to the sources in each category and to access details of them via the study's online spreadsheet (https://goo.gl/Z4jxBd).

### 4.2.3 RQ 2.3– Applicability scope/scale of the MMs

For RQ 2.3, we investigated the applicability scope of the MMs. As discussed in Section 3.5, the possible scopes could be one of the followings:

- a single department of a hospital/healthcare institution;
- multiple departments of a hospital;
- a single hospital/healthcare institution; multiple healthcare institutions;
- a city, region or province; government (nation-wide);
- international;
- or "other".

Based on the classification of the sources to the above categories, Figure 7 shows the histogram. We remind that, in eliciting the applicability scope of the MMs, we considered the expected/possible applicability of the models, and not necessarily the scale of the actual empirical study presented in a given source.

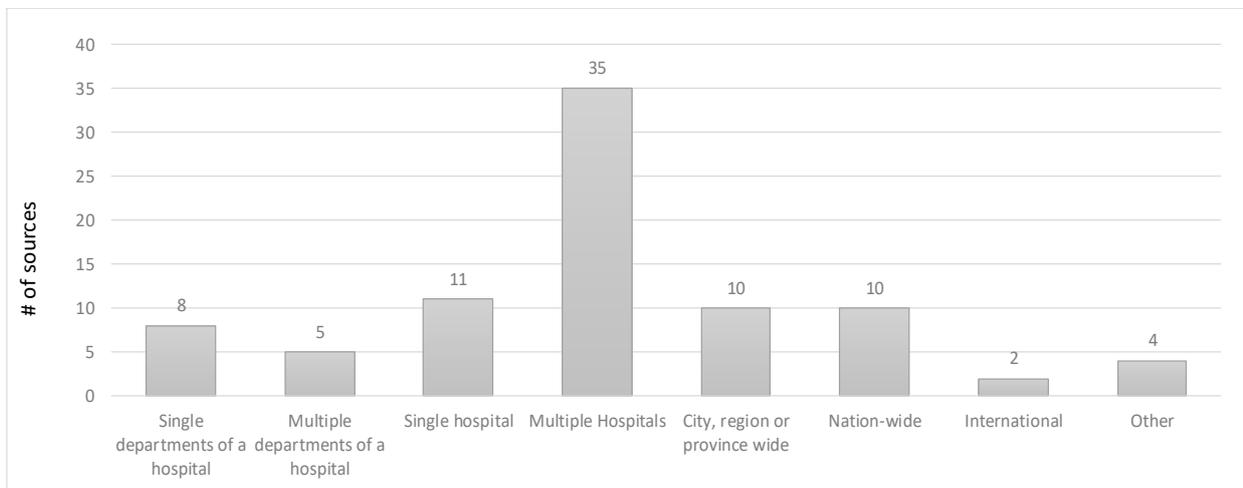

| | |
|---:|:---|
| Single departments of a hospital : | [Sources 4, 11, 17, 31, 37, 68, 80, 88] |
| Multiple departments of a hospital : | [Sources 37, 71, 74, 80, 85] |
| Single hospital : | [Sources 2, 3, 17, 19, 20, 36, 37, 56, 59, 80] |
| Multiple hospitals : | [Sources 4, 12, 15, 24, 26, 32, 35, 37, 39, 41, 42, 46, 49, 54, 55, 57, 61, 63, 64, 71, 74, 75, 76, 77, 78, 79, 80, 83, 86, 93, 94, 97, 100, 101] |
| City, region or province wide : | [Sources 1, 18, 22, 25, 37, 48, 70, 72, 93, 94] |
| Nation-wide : | [Sources 15, 30, 34, 37, 40, 58, 62, 93, 94, 95] |
| International : | [Sources 3, 81] |
| Other : | [Sources 53, 65, 82, 84, 9] |

**Figure 7- Sources with respect to the applicability scope/scale of MMs**

We found that the most common application scope/scale of MMs is "multiple hospitals" as addressed by 35 sources. For example, [Source 4] applied a maturity model for care pathways to 11 hospitals in the Netherlands. To enable "precision medicine" [44], [Source 24] measured the maturity of Business Intelligence (BI) in six hospitals in Italy. As another example, [Source 75] assessed relationship of organisational maturity and project success in healthcare based on the data from seven hospitals in Portugal.

## 4.3 RQ 3– PROPERTIES OF EMPIRICAL STUDIES

### 4.3.1 RQ 3.1– Scales of empirical studies

To characterize the scale of reported empirical studies, we focused on the number of healthcare institutions that a proposed MM was applied in. We also looked into other attributes of the empirical context, for example, the number of physicians or other types of experts involved in the surveys, or regular participants who were interviewed, or the number of workshops, focus groups, or expert panels that were organized to gather data for the maturity assessment. Note that the applicability scope/scale in the RQ 2.3 differs from the empirical scale in this RQ in that, in the case of the former, we assessed the "potential" applicability, while in the latter we looked at the "actual" reported empirical results presented in a source.

Out of the 101 sources, 56 reported quantitative information regarding the scale of their empirical studies. There are 33 sources that performed their applications in healthcare institutions (hospitals, medical centers, or clinics), while the remaining 23 sources performed empirical works based on other forms, such as opinion surveys, workshops or focus groups with healthcare specialists, or interviews with experts. In the last group ("other" forms), 7 sources used surveys with diverse stakeholders in the healthcare domain (e.g., [Sources 70, 71, 84, 88]). Five other studies applied focus groups/expert panels for the evaluation of the MMs (e.g., [Sources 76 and 77]).

These sources typically structured their evaluations as case studies, and involved experts from multiple departments within an institution (e.g., [Sources 2 and 19]). The sources that performed their empirical evaluations in multiple institutions (22 sources), on the other hand, varied greatly in terms of the number of institutions. There were 4 sources that applied their MMs in large scale, i.e., above 50 institutions. [Source 35] and [Source 78] involved the maximum numbers of institutions in our list of sources, in 538 and 129 hospitals, respectively. Sources that involved high number of institutions in their research typically conducted opinion surveys, which incorporated questions from maturity models to gain an understanding of the state-of-the-maturity of healthcare in particular geographical regions. These include, for instance, the maturity of quality improvement systems across Europe [Source 32], e-healthcare maturity in hospitals in Taiwan [Source 35], or process management capability of hospitals in Switzerland [Source 78].

It is also important to note that a large majority of sources that reported quantitative data regarding the scale of their empirical studies originated from academia (52 out of 55 sources). Only 3 sources [Source 84, 88, and 97] in the grey literature (out of 21 in total) provided quantitative information about the empirical studies that were conducted.

**4.3.2 RQ 3.2 - Empirical benefits of applying MMs**

In addressing RQ 3.2, we extracted and synthesized the benefits reported by sources as the results of their empirical studies. In doing so, we focused particularly on the benefits reported through the application of the MMs, and whether these benefits were stated in quantitative or qualitative forms (or both). We observed that more than half of the sources (62 out of 101) reported the benefits of applying MMs in qualitative terms, and only 2 sources reported benefits in quantitative terms [Sources 70, 79]. The remaining 37 sources did not provide any explicit discussions on the benefits of applying MMs.

As an example to quantitative benefits, [Source 70] measured maturity of use for electronic medical records (EMRs). Comparing the quantitative data from before and after applying the maturity assessment showed an increase from 21% to 83% of physicians who would using EMRs as the principal method of record-keeping. As another example, [Source 79] compared procurement maturity with procurement performance in three Dutch hospitals. The gathered numerical measurements supported the hypothesis that an increase in maturity of organization, processes and IT in hospitals would lead to better procurement performance, thus again showing a quantitative benefit of the proposed MM.

We then looked at the 62 sources, which have reported the benefits of applying MMs in qualitative terms. To better understand the vast amount of qualitative data that we extracted from the papers, we classified them using the "open coding" (grounded theory) method [45] which are widely used for qualitative analysis. The process resulted in three classifications: (1) Identifying issues and providing guidance for improvement (n=17); (2) Improving efficiency, effectiveness, performance, and productivity (n=12); and (3) "Other" benefits (n=33). The classifications of the sources that report these types of benefits, together with selected examples, are shown in Table 5.

Based on these numbers, we can see that although the number of sources that reported on the benefits of applying MMs in concrete quantitative terms was limited, there was a vast number of sources that reported on the benefits of using MMs in healthcare in qualitative terms.

Table 5- Qualitative benefits of applying MMs

| Qualitative benefits | # of sources | References to the sources | Example qualitative benefit |
|---|---|---|---|
| Identifying issues and providing guidance for improvement | 17 | [Sources 1, 3, 4, 17, 19, 22, 30, 36, 41, 46, 55, 68,74, 85, 87, 92, 101] | "The employees noted that the capabilities included in the model were highly relevant to their situation and provide plenty of discussion points in regards to areas for improvement." [Source 3] <br> "The results indicate the usefulness of the proposed model in assessing pathway's maturity and its potential to provide guidance for its improvement." [Source 4] |

| | | | "The MM provides support for identifying deficiencies in strategic, organizational, and technical cooperation capabilities." [Source 46] |
|---|---|---|---|
| Improving efficiency, effectiveness, performance, and productivity | 12 | [Sources 18, 21, 26, 39, 60, 63, 81, 83, 84, 95, 97, 99] | "The MMs allow healthcare organisations systematically improve performance." [Source 39]<br>"The MMs could enable hospital to deliver benefits quicker and more economically." [Source 95] |
| "Other" benefits | 33 | [Sources 6, 11, 12, 14, 21, 22, 23, 24, 31, 41, 43, 44, 48, 51, 52, 53, 56, 60, 63, 68, 73, 74, 78, 79, 80, 81, 82, 84, 85, 87, 88, 96, 100] | "Sustaining healthcare delivery" [Source 6]<br>"Allowing a quick and reproducible analysis of a (healthcare) center within a few days" [Source 11]<br>"More ready access to healthcare" [Source 23]<br>"Providing useful elements for developing an action plan" [Source 24]<br>"A means of internal and external benchmarking, self-assessment, change management, and organisational learning" [Source 31] |

## 5 DISCUSSION

### 5.1 OBSERVATIONS AND SUGGESTIONS

In response to RQ1, we reviewed sources by contribution and research types. The general observable trend showed that new MMs are continuously being offered in this area and there were 68 new MMs proposed. Also, the trend of sources with emphasis on and having major contributions as empirical (case) studies was an indicator of progressing nature of MM adoption. In terms of research-method types, a large ratio of sources fell under the categories of "weak empirical study" and "solution proposal". Although this is noted as a good sign of emergence and empiricism in this area, there is a need for studies with stronger empirical evidence.

For RQ 2, we reviewed sources by various properties of MMs, including; the identity of the proposed models, aspects covered, and applicability scope/scale. We identified and listed MMs proposed by the sources, and also extracted and indexed the subject-areas that they focused on. We believe the catalogue describing the available MMs and the index given in [35] could be useful to practitioners, decision makers, and researchers who are interested to assess maturity of a given healthcare entity (e.g., a hospital) from a particular point of view.

The list of subject-focuses demonstrated a wide range of research and application, which is promising for further research and practice. Since the index of subject areas is very broad, detailed analysis of critical areas for potential future work demands for various specialties in healthcare and technical domains. Therefore, we leave this duty for future research. As an example, we highlight areas such as practical aspects of patient-safety (e.g. incorrect labeling in patients' drug charts), and the clinical safety in medical and surgical departments. We were able to locate and review three MMs related to patient safety: (1) patient safety culture improvement tool [Source 76], (2) Manchester patient safety framework [Source 77], and (3) safety culture assessment in community [Sources 19, 85]. While these models were useful, we see a need for improvement to cover a broader range of clinical safety.

The MMs also varied in the aspects they cover, e.g., technology, business process, and people. The majority of the models address both "process" and "technology" aspects, and about half of the models cover "people" aspect. Almost one-third of the models widen their coverage to include "other" aspects such as culture, strategy, governance, leadership, interoperability, and data. As the included aspects demonstrated, this area is multi-disciplinary in nature and requires organization of research and practice in such challenging settings. We believe that the increasing amount of digitalization in healthcare will make the advantages of using guiding models as MMs even more significant in the upcoming years.

According to applicability scope/scale of MMs, the majority of the models are targeted for use in "multiple hospitals", followed by the scopes of "a single hospital", "city, region, or province wide", "nation-wide", "single departments of a hospital", and "multiple departments of a hospital". As a result, there is a balanced variety in applicability scale of the models. We suggest that the use of MMs among multiple hospitals or within governmental bodies (i.e. nation-wide) will be useful especially for benchmarking of healthcare services.

RQ 3 focused on the application scale of the empirical studies performed and the benefits observed. Findings on the empirical applications of MMs showed that there was a variety in the number of institutions involved and in the empirical research methods employed, e.g., case studies, surveys, interviews, workshops/focus groups. More than half of the sources reported on the scale of their works, and more than half of those reporting information about the scale of application performed their work directly in one or more healthcare institutions. More specifically, the MMs were empirically applied in various contexts from a single hospital to multiple hospitals and to health bodies of countries. Therefore, we argue that an important number of sources rely on a solid empirical basis. On the other hand, considerable percentage of works, particularly in the grey literature, lack application or empirical results.

Regarding the benefits of empirical applications, many of the sources reported on the usefulness and utility of the models through qualitative methods, but few presented concrete evidence on the benefits in using them. These benefits included, for instance, increased effectiveness and efficiency in healthcare service provisioning, improved patient satisfaction, and improved service quality. Moreover, there is a need to investigate the relation between qualitative and quantitative benefits in future studies. Designing longitudinal studies to evaluate the effects of increasing maturity in combination with understanding these effects in quantitative terms, e.g., by using process analytics techniques such as process mining [46-48], might be useful for this purpose.

## 5.2 THREATS TO VALIDITY

We discuss below the potential validity threats in the context of the four types adopted from [49].

*Internal validity*: In the systematic approach utilized for source selection; search engines, search terms and inclusion/exclusion criteria were carefully defined to ensure that this review is repeatable. Still, the selection of search terms and search engines, and bias in applying exclusion/inclusion criteria can be considered as the limitations during this process. In order to mitigate the risk of missing some relevant sources, formal searching using defined keywords was conducted, and an adequate and inclusive basis has been collected while building the study pool. In order to minimize the bias related to researchers' judgment and experience in applying inclusion/exclusion criteria, joint voting was applied after the initial source inclusion and only sources passing the joint voting were selected for review.

*Construct validity*: This type of validity is concerned with suitability of RQs and categorization scheme used for the data extraction. Review questions were designed to cover our goal, and answered according to a categorization scheme. For designing a good categorization scheme, we have adapted standard classifications and also have finalized the schema through several iterations. Another threat comes from the lack of empirical evidence in the primary sources. The majority of the grey literature was opinion or experience based and as the source of knowledge was not typically revealed, we are faced with an epistemological problem, i.e., we do not know, how we know, what we know. However, handling this limitation in study design by excluding the grey literature would leave an important source and related voice of practice out of the scope. Therefore, we decided to include these sources as they are in order to have a more complete profile in our review.

*Conclusion validity*: This type of validity requires reaching appropriate conclusions through rigorous and repeatable treatment. In order to ensure reliability of our treatments, the entire pool of primary sources was analyzed and the data was reviewed, extracted and synthesized by the authors. Following the systematic approach and describing review questions and procedure ensured replicability of this study and created confidence in that the results of a similar study would not have major deviations from our classification.

*External validity*: As described earlier, search terms in the source selection approach resulted in having primary sources all written in English language. However, good proportion of industrial and collaborative works, in addition to academic studies, exists in our sources due to including the grey literature. This means that our inclusive process of article selection has lead us to have an adequate basis for concluding results and that our pool contained sufficient information to represent the knowledge reported by other researchers and professionals. Nevertheless, the findings of this study are mainly within the specific area under study, and we have no intention to generalize our results beyond this field.

## 6 CONCLUSION

Complex and multi-disciplinary nature of healthcare domain brings significant challenges in terms of process and people management, and information technology and systems. Moreover, increasing amount of digitalization of healthcare services demands for well-defined guidelines and success reports to rationalize and effectively manage the transition process. In response to those challenges, a wide variety of maturity assessment and maturity models has been proposed and applied in the domain.

In this study, we provided an overview of these models and classified them based on their various properties and applications using MLR as the research method. The results of our review indicated that MMs are continuously being offered under various subjects of healthcare and also empirically applied, and that the body of sources presents a good balance between research and practice, which is promising to conduct further studies in this area. Future research, however, is required to include results from a stronger empirical basis, especially for the reports in the grey literature.

The index of the subject areas concerned by the MMs is so broad that there is a need for deeper analysis of the sources with respect to detailed specialties of healthcare and technology. It is suggested that the researchers should check the list of subject focuses by the MMs that we reported in this study, prior to proposing new models. This is especially important for eliminating identical studies (i.e. reinventing the wheel) as well as for creating opportunity on using or adapting already available MMs.

Finally, we believe that conducting longitudinal studies in different contexts by using quantitative or mixed methods (e.g. process mining) could better demonstrate the advantages of using MMs and therefore serve the community with stronger empirical evidence to advocate the utility of these models – particularly on the side of the practitioners.

## REFERENCES

[1]  S. Quaglini, "Information and communication technology for process management in healthcare: a contribution to change the culture of blame," *Journal of Software Maintenance and Evolution: Research and Practice,* vol. 22, no. 6-7, pp. 435-448, 2010.

[2]  M. Kirchmer, S. Laengle, and V. Masías, "Transparency-Driven Business Process Management in Healthcare Settings [Leading Edge]," *IEEE Technology and Society Magazine,* vol. 32, no. 4, pp. 14-16, 2013.

[3]  A. Tarhan, O. Turetken, and H. Reijers, "Do Mature Business Processes Lead to Improved Performance? - A Review of Literature for Empirical Evidence," in *European Conference on Information Systems*, 2015.

[4]  P. Crosby, *Quality is Free*. McGraw-Hill, 1979.

[5]  A. Stevenson and M. Waite, *Oxford English Dictionary – The Definitive Record of the English Language*. Oxford University Publisher, 2004.

[6]  J. Becker, R. Knackstedt, and J. Poeppelbuss, "Developing Maturity Models for IT Management - A Procedure Model and its Application," vol. 1, no. 3, pp. 213-222, 2009.

[7]  A. Tarhan, O. Turetken, and H. A. Reijers, "Business process maturity models: A systematic literature review," *Information and Software Technology,* vol. 75, pp. 122-134, 2016.

[8]  OMG, "Business Process Maturity Model (BPMM), Ver.1," 2008.

[9]  *ISO/IEC 15504-5: Information Technology-Process Assessment. Part–5: An exemplar software life cycle process assessment model*, 2012.

[10] M. C. Paulk, *The Capability Maturity Model: Guidelines for Improving the Software Process*. Addison-Wesley, 1995.

[11] J. V. Carvalho, Á. Rocha, and A. Abreu, "Maturity Models of Healthcare Information Systems and Technologies: a Literature Review," *Journal of Medical Systems,* journal article vol. 40, no. 6, pp. 1-10, 2016.

[12] M. Söylemez and A. Tarhan, "The Use of Maturity/Capability Frameworks for Healthcare Process Assessment and Improvement," in *Software Process Improvement and Capability Determination: 16th International Conference, SPICE 2016, Dublin, Ireland, June 9-10, 2016, Proceedings*, M. P. Clarke, V. R. O'Connor, T. Rout, and A. Dorling, Eds. Cham: Springer International Publishing, 2016, pp. 31-42.

[13] R. T. Ogawa and B. Malen, "Towards Rigor in Reviews of Multivocal Literatures: Applying the Exploratory Case Study Method," *Review of Educational Research,* vol. 61, no. 3, pp. 265-286, 1991.

[14] L. McAuley, B. Pham, P. Tugwell, and D. Moher, "Does the inclusion of grey literature influence estimates of intervention effectiveness reported in meta-analyses?," *The Lancet,* vol. 356, no. 9237, pp. 1228-1231, 2000.

[15] V. Alberani, P. De Castro Pietrangeli, and A. M. Mazza, "The use of grey literature in health sciences: a preliminary survey," *Bulletin of the Medical Library Association,* vol. 78, no. 4, pp. 358-363, 1990.

[16] K. M. Benzies, S. Premji, K. A. Hayden, and K. Serrett, "State-of-the-Evidence Reviews: Advantages and Challenges of Including Grey Literature," *Worldviews on Evidence-Based Nursing,* vol. 3, no. 2, pp. 55-61, 2006.

[17] V. Garousi, M. Felderer, and M. V. Mäntylä, "Guidelines for conducting multivocal literature reviews in software engineering," *Information and Software Technology,* vol. 106, pp. 101-121, 2019.

[18] R. van de Wetering and R. Batenburg, "A PACS maturity model: A systematic meta-analytic review on maturation and evolvability of PACS in the hospital enterprise," *International Journal of Medical Informatics,* vol. 78, no. 2, pp. 127-140, 2009.

[19] R. Álvaro, "Evolution of Information Systems and Technologies Maturity in Healthcare," *International Journal of Healthcare Information Systems and Informatics (IJHISI),* vol. 6, no. 2, pp. 28-36, 2011.

[20] E. Joukes, N. de Keizer, and R. Cornet, "Composite quality of care scores, electronic health record maturity models, and their associations; preliminary literature review results," *Stud Health Technol Inform,* vol. 192, p. 981, 2013.

[21] P. Brooks, O. El-Gayar, and S. Sarnikar, "Towards a Business Intelligence Maturity Model for Healthcare," in *2013 46th Hawaii International Conference on System Sciences*, 2013, pp. 3807-3816.

[22] M. D. I. C. (MDIC);, "Quality System Maturity Model for Medical Devices-Medical Device Innovation Consortium (MDIC)," *http://mdic.org/wp-content/uploads/2015/06/MDIC-Deloitte_Maturity-Model-Research-Report_June-2015.pdf,* Last accessed: Oct 1, 2019.

[23] K. Flott, R. Callahan, A. Darzi, and E. Mayer, "A Patient-Centered Framework for Evaluating Digital Maturity of Health Services: A Systematic Review," *J Med Internet Res,* vol. 18, no. 4, p. e75, 2016.

[24] J. V. Carvalho, Á. Rocha, and A. Abreu, "Information Systems and Technologies Maturity Models for Healthcare: A Systematic Literature Review," in *New Advances in Information Systems and Technologies: Volume 2*, Á. Rocha, M. A. Correia, H. Adeli, P. L. Reis, and M. Mendonça Teixeira, Eds. Cham: Springer International Publishing, 2016, pp. 83-94.

[25] J. V. de Carvalho, Á. Rocha, and J. B. de Vasconcelos, "Maturity Models for Hospital Information Systems Management: Are They Mature?," in *Innovation in Medicine and Healthcare 2015*, Y.-W. Chen, C. Torro, S. Tanaka, J. R. Howlett, and L. C. Jain, Eds. Cham: Springer International Publishing, 2016, pp. 541-552.

[26] M. Söylemez and A. Tarhan, "A Review and Comparison of Maturity/Capability Frameworks for Healthcare Process Assessment and Improvement," *Software Quality Professional,* vol. 19, no. 2, 2017.

[27] J. V. Carvalho, Á. Rocha, R. van de Wetering, and A. Abreu, "A Maturity model for hospital information systems," *Journal of Business Research, In press,* 2017.

[28] M. W. van Tulder, W. J. J. Assendelft, B. W. Koes, and L. M. Bouter, "Method Guidelines for Systematic Reviews in the Cochrane Collaboration Back Review Group for Spinal Disorders," *Spine,* vol. 22, no. 20, pp. 2323-2330, 1997.

[29] J. Garrard, *Health Sciences Literature Review Made Easy*. Jones & Bartlett Publishers, 2016.

[30] M. E. Falagas, E. I. Pitsouni, G. A. Malietzis, and G. Pappas, "Comparison of PubMed, Scopus, Web of Science, and Google Scholar: strengths and weaknesses," *The Federation of American Societies for Experimental Biology Journal,* vol. 22, no. 2, pp. 338-342, 2008.

[31] M. Shultz, "Comparing test searches in PubMed and Google Scholar," *Journal of the Medical Library Association,* vol. 95, no. 4, pp. 442-445, 2007.

[32] P. Jacso, "As we may search: Comparison of major features of the Web of Science, Scopus, and Google Scholar citation-based and citation-enhanced databases," *Current Science,* vol. 89, no. 9, pp. 1537-1547, 2005.

[33] D. Giustini and E. Barsky, "A look at Google Scholar, PubMed, and Scirus: comparisons and recommendations," *Journal of the Canadian Health Libraries Association* vol. 26, no. 3, pp. 85-89, 2005.

[34] G. R. Samadzadeh, T. Rigi, and A. R. Ganjali, "Comparison of Four Search Engines and their efficacy With Emphasis on Literature Research in Addiction (Prevention and Treatment)," *International Journal of High Risk Behaviors & Addiction,* vol. 1, no. 4, pp. 166-171, 2013.

[35] A. Tarhan, V. Garousi, O. Turetken, M. Söylemez, and S. Garossi, "Dataset for paper: Maturity assessment and maturity models in healthcare: a multivocal literature review," *http://doi.org/10.5281/zenodo.2600722*, 2019.

[36] eBizMBA Inc., "Top 15 Most Popular Search Engines | July 2017," *http://www.ebizmba.com/articles/search-engines*, Last accessed: Oct 1, 2019.

[37] Y. McGrath, H. Sumnall, K. Edmonds, J. McVeigh, and M. Bellis, "Review of grey literature on drug prevention among young people," *National Institute for Health and Clinical Excellence,* 2006.

[38] J. Adams *et al.*, "Searching and synthesising 'grey literature' and 'grey information' in public health: critical reflections on three case studies," *Systematic Reviews*, journal article vol. 5, no. 1, p. 164, 2016.

[39] B. Kitchenham and S. Charters, "Guidelines for Performing Systematic Literature Reviews in Software engineering," in Evidence-Based Software Engineering," *Evidence-Based Software Engineering,* 2007.

[40] L. Gastaldi *et al.*, "Measuring the maturity of business intelligence in healthcare: Supporting the development of a roadmap toward precision medicine within ISMETT hospital," *Technological Forecasting and Social Change,* vol. 128, pp. 84-103, 2018.

[41] P. Bernard, *COBIT® 5 - A Management Guide*. Van Haren, 2012.

[42] C. Weber, B. Curtis, and M. B. Chrissis, "The capability maturity model, guidelines for improving the software process," *Harlow: Addison Wesley*, 1994.

[43] J. Lee, D. Lee, and S. Kang, "An Overview of the Business Process Maturity Model (BPMM)," Berlin, Heidelberg, 2007, pp. 384-395.

[44] H.-P. Deigner and M. Kohl, *Precision Medicine: Tools and Quantitative Approaches*. Academic Press, 2018.

[45] M. B. Miles, A. M. Huberman, and J. Saldana, *Qualitative Data Analysis: A Methods Sourcebook*, Third Edition ed. SAGE Publications Inc., 2014.

[46] D. Van, "Process Mining Discovery, Conformance and Enhancement of Business Processes," ed: Springer Heidelberg Dordrecht London New York. ISBN, 2011.

[47] T. G. Erdogan and A. Tarhan, "Systematic Mapping of Process Mining Studies in Healthcare," *IEEE ACCESS,* vol. 6, pp. 24543-24567, 2018.

[48] T. Gurgen Erdogan and A. Tarhan, "A Goal-Driven Evaluation Method Based On Process Mining for Healthcare Processes," *Applied Sciences,* vol. 8, no. 6, p. 894, 2018.

[49] C. Wohlin, P. Runeson, M. Höst, M. C. Ohlsson, B. Regnell, and A. Wesslén, *Experimentation in software engineering*. Springer Science & Business Media, 2012.